\documentclass[12pt]{iopart}  
\usepackage{graphicx}
\usepackage{mcite}
\usepackage{booktabs}
\usepackage{multirow}
\usepackage{xcolor}
\usepackage{epstopdf}

\begin{document}

\title[]{Two superconducting thin films systems with potential integration of different quantum functionalities}

\author{Snehal Mandal$^1$, Biplab Biswas$^1$, Suvankar Purakait$^2$, Anupam Roy$^3$, Biswarup Satpati$^2$, Indranil Das$^2$ and B. N. Dev$^1$}

\address{$^1$ Centre for Quantum Engineering, Research and Education, TCG-CREST,\\
EM Block, Sector-V, Salt Lake, Kolkata 700091, India.}
\address{$^2$ Saha Institute of Nuclear Physics, 1/AF, Bidhannagar, Kolkata 700064, India.}
\address{$^3$ Dept. of Physics, Birla Institute of Technology Mesra, Ranchi, Jharkhand 835215, India.}
\ead{bhupen.dev@gmail.com}
\vspace{10pt}

\begin{abstract}
Quantum computation based on superconducting circuits utilizes superconducting qubits with Josephson tunnel junctions. Engineering high-coherence qubits requires materials optimization. In this work, we present two superconducting thin film systems, grown on silicon (Si), and one obtained from the other via annealing. Cobalt (Co) thin films grown on Si were found to be superconducting [EPL 131 (2020) 47001]. These films also happen to be a self-organised hybrid superconductor/ferromagnet/superconductor (S/F/S) structure. The S/F/S hybrids are important for superconducting $\pi$-qubits [PRL 95 (2005) 097001] and in quantum information processing. Here we present our results on the superconductivity of a hybrid Co film followed by the superconductivity of a CoSi$_2$ film, which was prepared by annealing the Co film. CoSi$_2$, with its $1/f$ noise about three orders of magnitude smaller compared to the most commonly used superconductor aluminium (Al), is a promising material for high-coherence qubits. The hybrid Co film revealed superconducting transition temperature $T_c$ = 5 K and anisotropy in the upper critical field between the in-plane and out-of-plane directions. The anisotropy was of the order of ratio of lateral dimensions to thickness of the superconducting Co grains, suggesting a quasi-2D nature of superconductivity. On the other hand, CoSi$_2$ film showed a $T_c$ of 900 mK. In the resistivity vs. temperature curve, we observe a peak near $T_c$. Magnetic field scan as a function of $T$ shows a monotonic increase in intensity of this peak with temperature. The origin of the peak has been explained in terms of parallel resistive model for the particular measurement configuration. Although our CoSi$_2$ film contains grain boundaries, we observed a perpendicular critical field of 15 mT and a critical current density of 3.8$\times$10$^7$ A/m$^2$, comparable with epitaxial CoSi$_2$ films.
\end{abstract}
%
%

%
%
\ioptwocol

\section{Introduction}
Superconductivity plays a pivotal role in quantum technology, especially in quantum computers. The basic unit of a quantum computer is a quantum bit (or qubit), which can be implemented via several physical platforms. However, superconducting qubit is one of the most promising approaches towards building a scalable fault-tolerant quantum computer \cite{kjaergaard2020superconducting, lau2022nisq}. Superconducting qubit technology uses several superconducting materials, such as Al, Nb, Ta etc. for the fabrication of qubits, capacitor pads and resonators. One of the major aims of materials research in this area is to develop superior qubits with longer decoherence time. The main component of a superconducting qubit is a Josephson junction (JJ), which is a superconductor/insulator/superconductor (S/I/S) heterostructure \cite{nakamura1999coherent}. Although, several superconductors are used for building superconducting quantum computers, Al is the most widely used superconducting material for the fabrication of qubits and quantum processors. Aluminum-based qubits use Al/AlO$_x$/Al heterostructures as JJs. Since the quantum states are intrinsically fragile, interactions of qubits with the environment result in various sources of noise which lead to decoherence. Suppressing decoherence, or increasing decoherence time, involves a synchronous optimization of both electromagnetic design and materials quality \cite{siddiqi2021engineering}. Besides the improvement of qubit design and microwave engineering, it is desirable to use superconductors with superior material properties. One important source of noise, responsible for qubit decoherence, is the $1/f$ noise \cite{siddiqi2021engineering} originating from the interfaces and surfaces of the materials and heterostructures used for fabricating the qubits. Recently it was shown that cobalt disilicide (CoSi$_2$), a superconductor with comparable superconducting transition temperature ($T_c$) to that of Al \cite{chiu2017ultralow}, has two to three orders of magnitude less $1/f$ noise compared to Al. It was conjectured that CoSi$_2$ films may provide superior qubits with longer decoherence time compared to Al \cite{chiu2017ultralow}.

CoSi$_2$ has already been used for decades as metallic contacts in semiconductor technology \cite{zhang2003metal, furukawa1983epitaxial}. CoSi$_2$ is usually produced by depositing thin films of cobalt (Co) on silicon (Si) followed by post-deposition annealing \cite{zhang2003metal, shi2000growth}, or by directly depositing Co on hot Si substrates \cite{mahato2012nanodot, tung1982growth}. Out of these two materials - Co and CoSi$_2$ - although CoSi$_2$ is a superconductor \cite{chiu2017ultralow, tsutsumi1997superconductivity}, Co was neither known nor expected to be a superconductor as cobalt is a ferromagnetic metal, and materials possessing long range magnetic order do not exhibit superconductivity \cite{kittel2018introduction}. In some cases, such ferromagnetic materials, for example iron (Fe), can show superconductivity under high pressure \cite{shimizu2001superconductivity}. However, bulk Co was not shown to be a superconductor under any condition. Very recently, superconductivity was discovered in Co thin films grown on Si \cite{banu2020inhomogeneous}. Normal Co has a hexagonal close packed (hcp) crystal structure and is ferromagnetic. In the thin films, grown on clean Si substrates, a high-density non-magnetic (HDNM) face-centered cubic (fcc) phase of Co was found to have grown \cite{banu2017evidence, banu2018high}. While normal Co is not superconducting, this HDNM phase of Co is superconducting \cite{banu2020inhomogeneous}. Earlier, theoretically it was predicted that at a high-density the fcc phase of Co would lose magnetism \cite{yoo2000new}, although no superconductivity was predicted. However, the theoretical work carried out together with the experimental discovery of superconductivity in Co shows that at certain densities of fcc Co, it becomes non-magnetic as well as superconducting, and a detailed phase diagram has been produced. This phase diagram shows the range of densities and strains in the thin films for which superconductivity would be observed \cite{banu2020inhomogeneous}. Another interesting aspect of these Co films is that the whole film is not superconducting. The films have a self-organised hybrid three-layer structure $-$ HDNM-Co/Normal-Co/HDNM-Co \cite{banu2017evidence, banu2018high}. While the normal-Co is a ferromagnet (F), the HDNM-Co layers are superconductors (S). Thus, the Co films have an S/F/S hybrid structure. Such S/F/S hybrid structures have superconducting-spintronic and other applications in quantum technology, such as in quantum information processing \cite{baek2014hybrid, bhatia2022aspects}. S/F/S structures have also shown 0-$\pi$ quantum phase transition \cite{pompeo2014thermodynamic}. An S/F/S structure forms a ferromagnetic JJ (or $\pi$-JJ), which can be used to fabricate a superconducting $\pi$-qubit with a long decoherence time. Such a superconducting $\pi$-qubit may be a superconducting ring with one $\pi$-JJ and one normal JJ (or 0-JJ) structures \cite{yamashita2005superconducting}. 0-$\pi$ qubits can be implemented in different ways. Recently, an intrinsically error-protected superconducting 0-$\pi$ qubit with a long decoherence time (1.6 ms) has been experimentally realized \cite{gyenis2021experimental}.  

Thus, both the superconducting systems, Co and CoSi$_2$, have potential applications in quantum technology. With the possibility of localized pulsed laser annealing of a Co film on Si to form CoSi$_2$ \cite{lee2006multiple}, realization of devices incorporating functionalities of both the superconductors $-$ Co and CoSi$_2$, on the same Si substrate can be envisaged.

In this article, we explore the conversion of one superconducting system (Co) to another superconductor (CoSi$_2$), achieved via annealing and compare their superconducting properties with a view to utilize the latter in superconducting quantum circuits.  

\section{\label{Experimental Methods}Experimental methods}
Co films were deposited using electron beam evaporation onto a (100)-oriented n-type Si substrate (resistivity $\sim$ 3 - 8.5 $\Omega$.cm) at room temperature under high vacuum. Prior to deposition, the Si substrate was cleaned in 1\% aq. HF solution to etch-off the native oxide (SiO$_x$) layer. Earlier investigations have shown that this method of growth produces polycrystalline Co films, which are also superconducting \cite{banu2020inhomogeneous, banu2017evidence}. To investigate the superconductivity in the Co film, electrical resistivity measurements were performed using conventional linear four-probe technique (with indium-silver solder contacts) under externally applied magnetic fields inside a commercial physical property measurement system (PPMS of M/s. Quantum Design) down to low temperature (2 K). To apply magnetic fields along different directions with respect to the film (sample) plane, the sample was rotated using a rotator-puck/sample-insert provided with the PPMS. This has allowed us to measure critical field anisotropy, which was not investigated in earlier studies \cite{banu2020inhomogeneous}.

Following that, the superconducting Co film was converted into a superconducting CoSi$_2$ film via annealing at 850 $^o$C under high vacuum (HV) conditions ($<$ 5$\times$ $10^{-6}$ mbar) for 1.5 hours. To investigate the superconductivity of the CoSi$_2$ film, resistivity measurements were carried out using van der Pauw (vdP) technique inside a dilution refrigerator that could cool down to 30 millikelvin. The contact pads for the vdP configuration were prepared by thermal evaporation of Ti/Au, patterned by lift-off photolithography. The magnetic field could only be applied in the out-of-plane direction with the available sample space configuration inside the dilution refrigerator.

X-ray characterizations of both the Co film and the CoSi$_2$ film were done at room temperature in a Rigaku SmartLab diffractometer with Cu-K$_{\alpha}$ x-rays ($\lambda$ = 1.5406 \AA). X-ray reflectivity (XRR) was performed on the Co film to determine its thickness, roughness and electron densitiy distribution (layer structure), while grazing incidence x-ray diffraction (GIXRD) was performed on both the Co and the CoSi$_2$ films to identify their phases. Moreover, cross-sectional high resolution transmission electron microscopy (HRTEM) was performed on the CoSi$_2$ film to investigate the thickness and crystal orientation of the film with respect to the substrate.
\section{Results and Discussions}
\subsection{Cobalt film}
\subsubsection{\label{Char_Co}Characterization}
\begin{figure}[h]
\centering
\includegraphics[width=0.89\columnwidth]{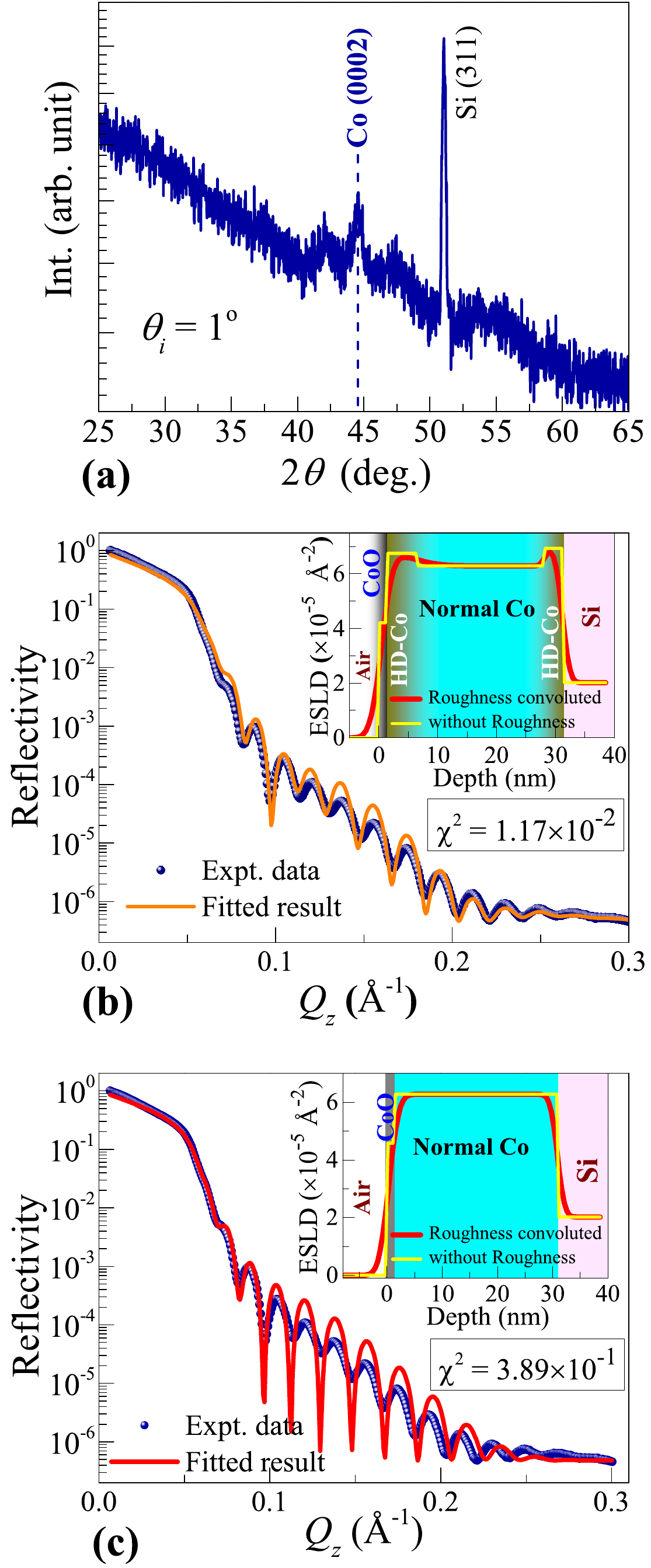}
\caption{\label{GIXRD_XRR_Co}(a) GIXRD pattern (at incident angle, $\theta_i$ = 1$^o$) of the Co film. The extra peak from the substrate was identified from $\phi$-scan measurement. (b) XRR of the Co film fitted with the tri-layer model as described in the text. (c) XRR of the same film fitted with Si/Normal-CO/CoO layer model, which shows relatively mediocre fit as compared to that in (b). Insets of both the panels show the ESLD profile corresponding to the XRR fits.}
\end{figure}

$ $\\
Figure \ref{GIXRD_XRR_Co} (a) shows GIXRD of the film, indicating the (0002) peak of Co. This indicates that the hcp phase of normal Co dominates the film. On the other hand, in order to get the information of the overall film stack (\textit{i.e.}, thickness, individual layer densities, surface and interfacial roughness), XRR data were carefully analyzed, since it can reveal the electron scattering length density (ESLD) or electron density depth profile (proportional to mass density depth profile) with very high depth resolution ($\sim$ 1 \AA), and rms roughness parameters in the direction normal to the film plane.
We obtained a total thickness of 30.89(3) nm for the Co film from the XRR measurement. The best fit to the XRR data ($\chi^2$ = 1.17$\times 10^{-2}$), as shown by solid orange-colored line in Fig. \ref{GIXRD_XRR_Co}(b), was obtained by considering a ESLD profile that shows
\begin{table}[h]
\caption{Depth dependent parameters of Co film extracted from XRR measurement:}
\label{XRR result}
\resizebox{\columnwidth}{!}{
\begin{tabular}{lccl}
\hline
Layer &
  \begin{tabular}[c]{@{}c@{}}Thickness\\ (nm)\end{tabular} &
  \begin{tabular}[c]{@{}c@{}}ESLD\\ ($\times$10$^{\textbf{-}5}$\AA$^{\textbf{-}2}$)\end{tabular} &
  \multicolumn{1}{c}{\begin{tabular}[c]{@{}c@{}}Roughness\\ (nm)\end{tabular}} \\ \hline
CoO          & 1.27(9)  & 4.53(7)  & 1.4(6) \\ \hline
HD-Co       & 1.61(7)  & 6.87(1)  & 1.2(9) \\ \hline
Normal Co    & 24.47(2) & 6.27(3)  & 1.0(0) \\ \hline
HD-Co       & 3.52(5)  & 6.70(5) & 0.5(5) \\ \hline
Si substrate & ---      & 2.01(5) & 1.0(3) \\ \hline
\end{tabular}
}
\end{table}
higher ESLDs at the Co/Si interface as well as near the Co/CoO interface (shown in the inset of the the panel) [A CoO layer ($\sim$2 nm) naturally grows on the Co film after removing it from the vacuum chamber following film deposition]. This kind of hybrid three-layer structure $-$ HDNM-Co/Normal-Co/HDNM-Co $-$ forms in a self-organised manner, similar to those reported in Ref. \cite{banu2017evidence, banu2018high}. The method of detailed analysis may be found in Refs. \cite{banu2017evidence, banu2018high}. The ESLD, thickness and roughness parameters obtained from XRR of the present sample are given in Table \ref{XRR result}. In our case, we obtained a layer of ESLD =  6.87(0) $\times$ 10$^{-5}$ \AA$^{-2}$ (of thickness 1.27(9) nm) close to the Co/CoO interface and another layer of ESLD = 6.70(5)$\times$ 10$^{-5}$ \AA$^{-2}$ (of thickness 3.52(5) nm) at the Co/Si interface, both of which are higher in density compared to that of Normal-Co layer (of thickness 24.47(2) nm) in the middle. [For comparison, we show that the XRR data could not be fitted properly assuming a single uniform density Normal Co layer (ESLD = 6.27(3) $\times$ 10$^{-5}$ \AA$^{-2}$) (with obvious CoO layer on top) along the whole depth of the film, as shown in Fig. \ref{GIXRD_XRR_Co} (c)]. Such HDNM Co layers in the obtained tri-layer structure are non-magnetic with nanoscale grains \cite{banu2017evidence, banu2018high} and show inhomogeneous superconductivity as reported in Ref. \cite{banu2020inhomogeneous}. Since the nature of superconductivity was reported to be inhomogeneous in nature, this stimulated us to investigate the superconducting behavior with variation of direction of magnetic field, as described below.
\subsubsection{\label{R-T_Co}Transport properties}
\begin{figure*}[h]
\centering
\includegraphics[width=0.95\textwidth]{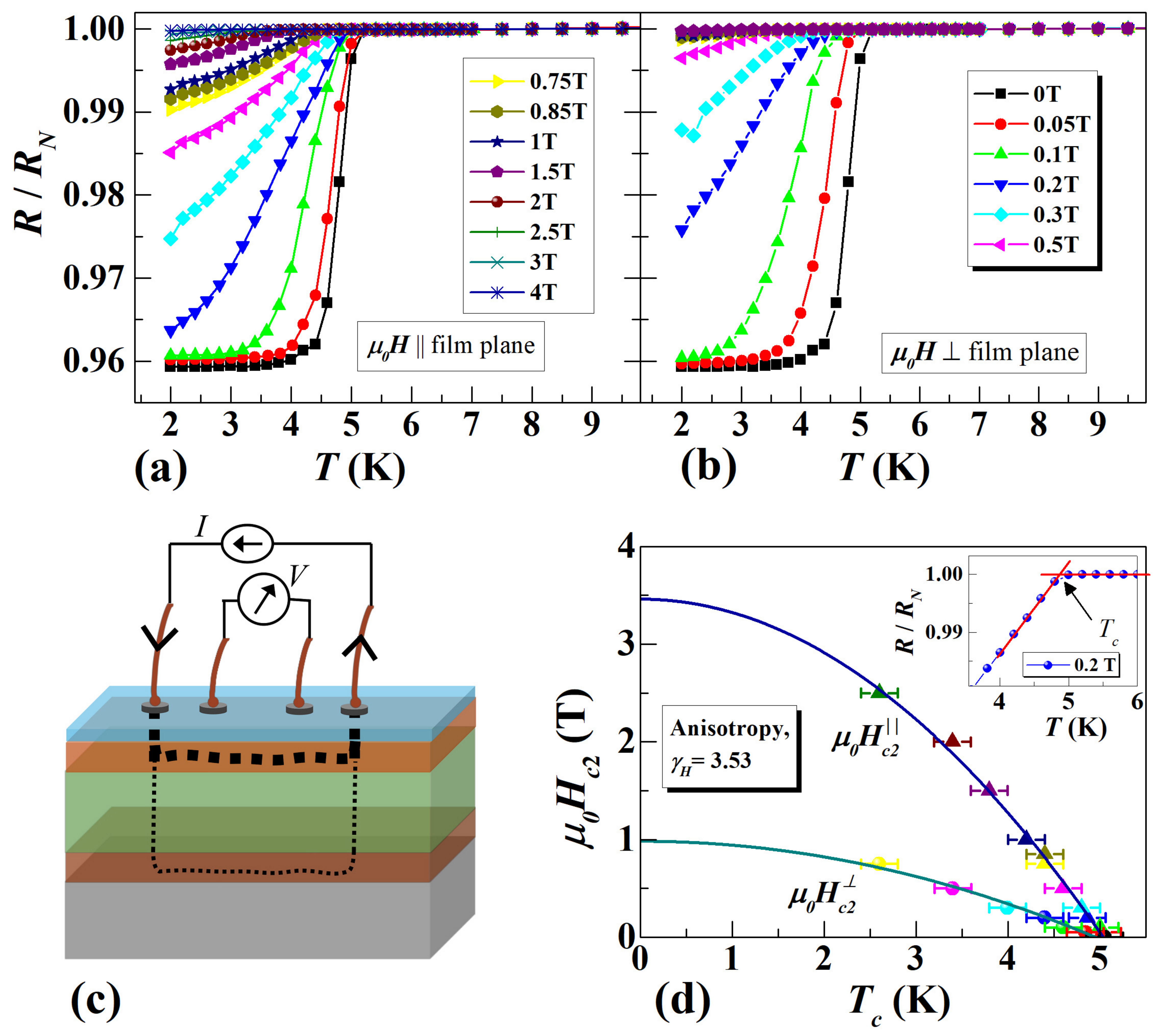}
\caption{\label{RT_PD_Co_fig} $R-T$ variation of the Co film at various magnetic fields applied: (a) along the film plane (in-plane/IP); (b) perpendicular to the film plane (out-of-plane/OOP). (c) Schematic of the linear four-probe measurement setup. (d)$\mu_0H_c - T_c$ phase diagram of the film showing clear anisotropy in the upper critical fields (Symbols: Data obtained from experimental $R-T$ curves, Solid lines: Fit using BCS model). Inset shows how we defined the $T_c$, marked by the arrow.}
\end{figure*}

$ $\\
To investigate the nature of superconductivity, we measure the temperature variation of resistance (down to $T$ = 2 K only) under various magnetic field strengths applied parallel (IP) and perpendicular (OOP) to the film plane. The results are presented in Fig. \ref{RT_PD_Co_fig}(a) and (b). We observe that the superconductivity is remarkably suppressed by a magnetic field of $\sim$ 2 T and $\sim$ 0.5 T applied IP and OOP, respectively.

However, one of the important features of these resistance ($R$) vs temperature ($T$) (or $R-T$) curves is the non-vanishing resistance in the superconducting state (\textit{i.e.}, below $T_c$, even in the absence of magnetic field). There can be three possibile contributions to this non-zero resistance: (i) due to presence of a comparatively higher resistive CoO layer at the top; (ii) due to contact resistance of the probes and sensitivity of the measuring instruments \cite{buckel2008superconductivity}; or (iii) due to granular nature of the sample itself \cite{adkins1980increased, durkin2020rare}. Moreover, our sample has superconducting Co layers in the background of a normal Co layer. In our sample, there is only 4$\%$ drop in resistance in the superconducting state with respect to the normal state. This can be explained considering the resistance contribution from the CoO layer and the normal Co layer [schematically shown in Fig. \ref{RT_PD_Co_fig}(c)], even if we neglect the contact resistance. It should be noted that the non-zero resistance in the superconducting state of granular metals might also be attributed to Josephson coupling between an increasing number of pairs of grains with decreasing temperature \cite{adkins1980increased, durkin2020rare, abeles1977effect, dynes1978two}.
By virtue of competition between the Josephson coupling energy ($E_{\rm{J}}$) and capacitive coupling energy ($E_c$) between the grains, the superconducting state appears when $E_{\rm{J}}$ $>>$ $E_c$.

To generate the $\mu_0H_c - T_c$ phase diagram, we first define the critical temperature ($T_c$) as the temperature where superconductivity sets in (onset of superconductivity) (as shown in the inset of Fig. \ref{RT_PD_Co_fig}(d) with an arrow). Thus we generate the phase diagram from the $R-T$ plots as shown in the main panel of Fig. \ref{RT_PD_Co_fig}(d), which reveals a clear anisotropy in the upper critical field along the two directions. We determine these upper critical fields namely $\mu_0H_{c2}^{||}$(0) and $\mu_0H_{c2}^{\perp}$(0) from fitting, using the BCS model:
\begin{equation}\label{BCS}
\mu_0H_{c2} = \mu_0H_{c2}(0) \left[ 1- \left( \frac{T}{T_{c0}} \right)^2\right]
\end{equation}
where $\mu_0H_{c2}$(0) is the zero temperature upper critical field and $T_{c0}$ is the absolute critical temperature. We also define the anisotropy in the upper critical field as:
\begin{equation}
\gamma_H = \frac{\mu_0H_{c2}^{||}(0)}{\mu_0H_{c2}^{\perp}(0)}
\end{equation}
\begin{table}[t]
\centering
\caption{Parameters extracted from the BCS fit to the $\mu_0H_c - T_c$ phase diagram:}
\label{Tab_Co}
\resizebox{\columnwidth}{!}{%
\begin{tabular}{ccccc}
\toprule
\begin{tabular}[c]{@{}c@{}}Field\\ Direction\end{tabular} & \begin{tabular}[c]{@{}c@{}}$\mu_0H_{c2}$(0)\\ (T)\end{tabular} & \begin{tabular}[c]{@{}c@{}}$T_c$(0)\\ (K)\end{tabular} & $\gamma_H$ & \begin{tabular}[c]{@{}c@{}}$\xi_0$\\ (nm)\end{tabular} \\ \midrule
IP ($\mu_0H_{c2}^{||}$) & 3.46(6) & 5.03(1) & \multirow{2}{*}{3.53} & 9.75(3) \\
OOP ($\mu_0H_{c2}^{\perp}$) & 0.98(4) & 4.94(5) &  & 18.32(2) \\ \bottomrule
\end{tabular}%
}
\end{table}
The obtained values are summarized in Table \ref{Tab_Co}. In comparison to robust 2D superconducting systems where anisotropy is quite strong ($\gamma_H$ $\sim$ 10), the anisotropy of our system ($\gamma_H$ $\approx$ 3.5) indicates a quasi-2D nature of superconductivity in it. We have layered superconducting regions (or grains) of HDNM Co whose aspect ratio of lateral dimension ($\approx$ 10 nm) to thickness ($\approx$ 3 nm) \cite{banu2017evidence} is almost 3. Therefore it is expected that the coherence length along the film-normal is longer than that along the film-plane, which induces an anisotropic pair-breaking mechanism depending on the direction of the magnetic field applied to the film. This explains anisotropy in the upper critical fields. 

In case of the orbital-limiting effect, which is dominant under perpendicular (OOP) magnetic fields, Cooper pair breaking is induced by the momentum, $e\vec{A}/\hbar c$, where $\vec{A}$ is the vector potential, and eventually, the kinetic energy of supercurrent exceeds the superconducting gap energy \cite{helfand1966temperature, werthamer1966temperature}. Thus, the upper critical field is recognized as the orbital-limiting field, $\mu_0H_{orb}$ = $\phi_0$/($2\pi \xi^2$), which depends on the coherence length of Cooper pair, $\xi$. Here $\phi_0$ is the magnetic flux quantum.

On the other hand, for a 2D weak coupling BCS superconductor, the upper critical field becomes limited by the so called Pauli paramagnetic limit (or the Chandrasekhar-Clogston limit) \cite{chandrasekhar1962note, clogston1962upper}, where the Zeeman splitting energy of individual electron spin exceeds the superconducting energy gap and thus Cooper pair becomes energetically unstable. This Pauli limiting field becomes important under parallel (IP) magnetic fields and is given by $\mu_0H_{\rm{P}}$ $\approx$ $\frac{1.76 k_{\rm{B}}T_c}{\sqrt{2}\mu_{\rm{B}}}$, where $k_{\rm{B}}$ and $\mu_{\rm{B}}$ are the Boltzmann’s constant and Bohr magneton, respectively. Taking $T_c$ = 5 K, we have $\mu_0H_{\rm{P}}$ = 9.2 T, which is way larger than the $\mu_0H_{c2}^{||}$(0) of our Co film suggesting a very weak coupling regime of superconductivity.

\subsection{CoSi$_2$ film}
We now turn towards the CoSi$_2$ film formed by annealing of the Co / Si thin film system discussed so far.
\begin{figure}[b]
\centering
\includegraphics[width=0.95\columnwidth]{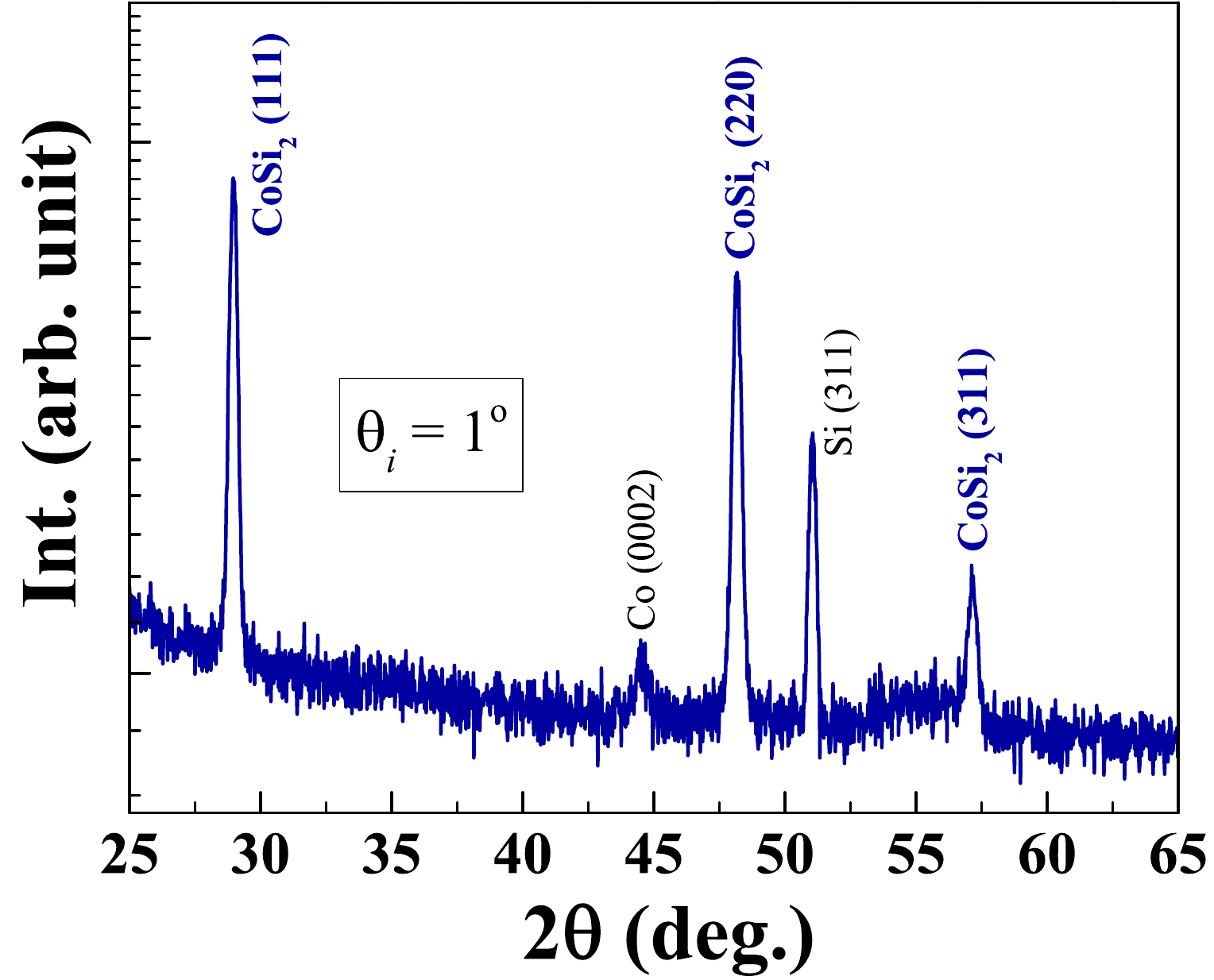}
\caption{\label{GIXRD_CoSi2}GIXRD pattern (at incident angle, $\theta_i$ = 1$^o$) of the CoSi$_2$ film. The extra peak from the substrate was identified from $\phi$-scan measurement.}
\end{figure}
\subsubsection{Characterization}$ $\\
\begin{figure*}[h]
\centering
\includegraphics[width=\textwidth]{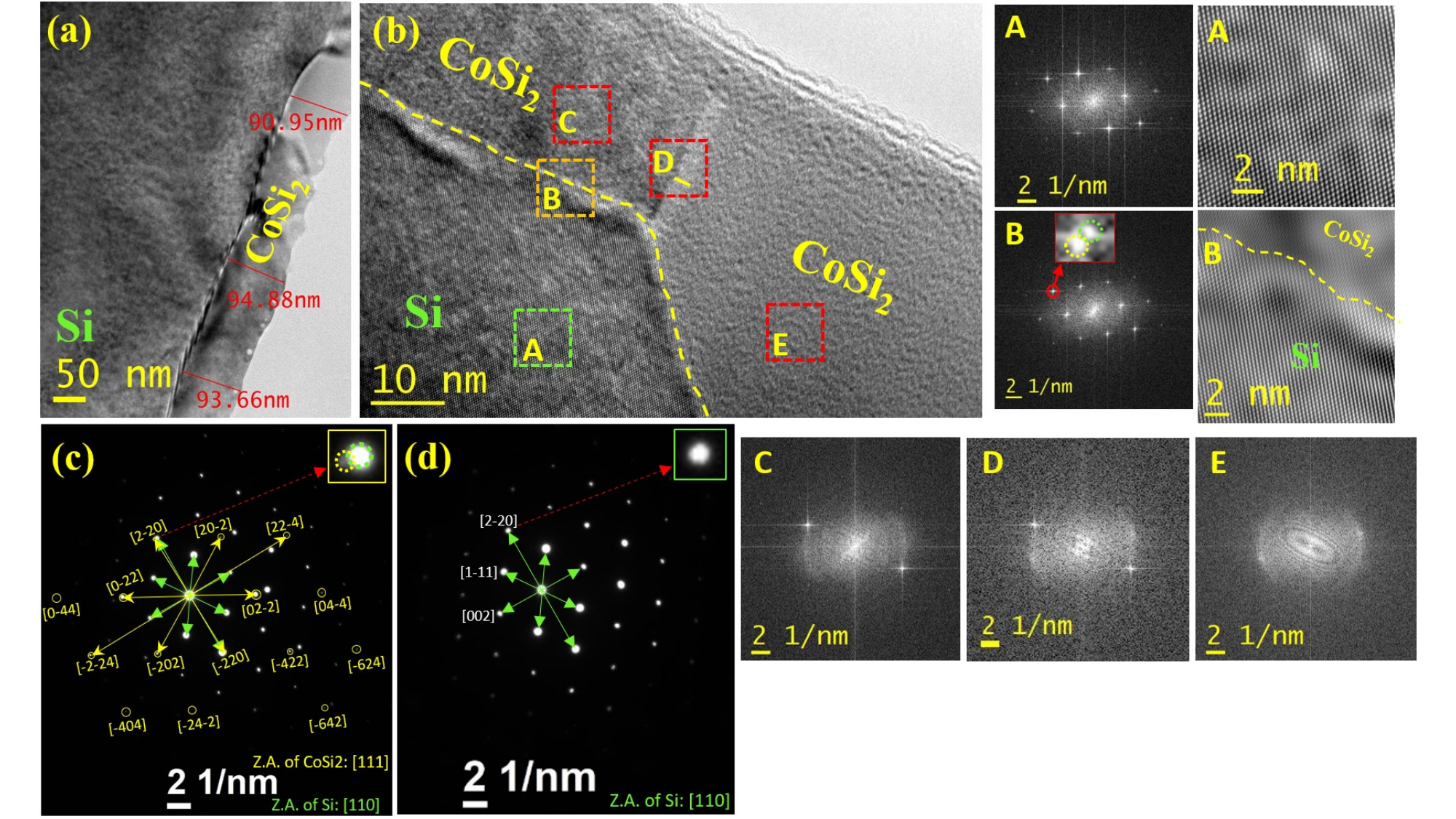}
\caption{\label{TEM} (a) Cross-sectional TEM image of the Si/CoSi$_2$ sample at low magnification. The average thickness of CoSi$_2$ is around 90 nm. (b) HRTEM image showing two different grains within the CoSi$_2$ film. FFT is performed on the regions marked A to E. The FFT image in B (substrate/film interface) shows splitting of the spot due to different crystal orientation of CoSi$_2$ with respect to the Si substrate. Also, the different orientation of the two grains of CoSi$_2$ is clear from the corresponding FFT images as we gradually go from region C, D (grain boundary region) to E. (c) SAED patterns of CoSi$_2$ film + Si substrate, showing splitting of the marked [2$\bar{2}$0] spot (magnified in the top corner of the same image). (d) SAED pattern of the Si substrate only, showing a single spot at the [2$\bar{2}$0] position. Note that the images were collected by setting zone-axis (Z.A.) parallel to [110] direction of Si. Analysis of the SAED patterns reveal that Z.A. of CoSi$_2$ is parallel its (111) direction, which means, orientation of CoSi$_2$ is such that (111) planes of CoSi$_2$ is parallel to (110) planes of Si.}
\end{figure*}
Figure \ref{GIXRD_CoSi2} shows the GIXRD pattern of the annealed sample. The peaks at 28.84$^o$, 48.12$^o$ and 57.18$^o$ correspond to the (111), (220), and (311) orientations of CoSi$_2$ and confirm that the CoSi$_2$ phase has already formed with 1.5 hours of annealing at 850 $^o$C and manifest good crystallinity of the film. To get further information about the crystalline nature of the film, the cross-sectional HRTEM of the film has been performed, as shown in Fig. \ref{TEM}(a)-(d). From the low magnification image [Fig. \ref{TEM}(a)], the overall film thickness is found to be around 90 nm. The high magnification HRTEM image in Fig. \ref{TEM}(b) further confirms that the film has multiple CoSi$_2$ grains. The relative orientation of the CoSi$_2$ film with respect to the Si substrate can be visualised from the SAED pattern in figure panel (c). For comparison the SAED pattern of the Si substrate is shown in panel (d). In this film, (111) planes of CoSi$_2$ is parallel to (110) planes of Si. To get further insight into the detailed crystallinity, the fast Fourier transform (FFT) has been performed in various regions as marked by (A) to (E) of the image in Fig. \ref{TEM}(b) and are shown in the corresponding sub-panels. It is interesting to note the changes in the FFT diffraction patterns as we go from one grain to the other across the grain boundary [sub-panels (C) to (E)]. This confirms the different orientations of the two grains separated by the grain boundary. It should be noted that two such superconducting grains, separated by a grain boundary, can behave as an inherent Josephson junction, the effects of which can be seen in the resistivity and $I-V$ characteristics of the sample.

\subsubsection{Transport Properties}$ $\\
\textit{A. Resistivity}\\
\begin{figure*}[ht]
\centering
\includegraphics[width=0.95\textwidth]{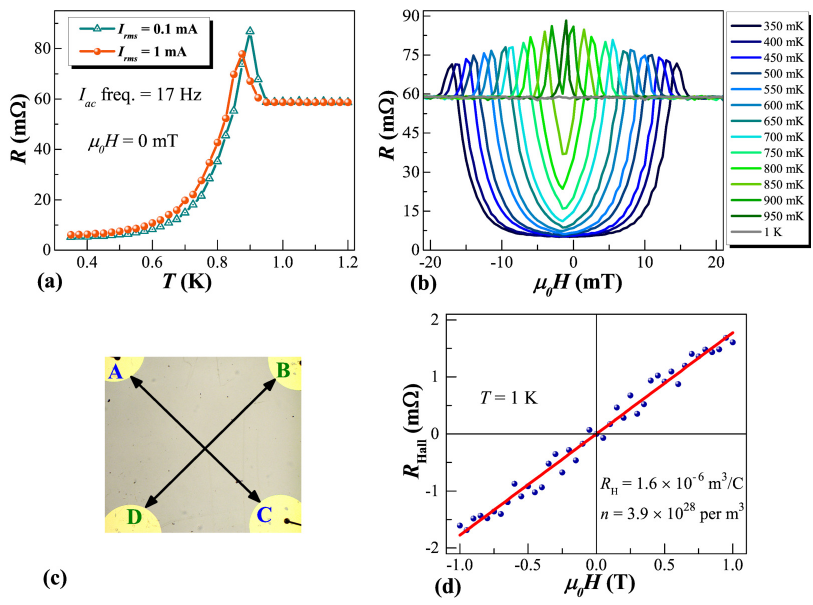}
\caption{\label{RT_RH_CoSi2} (a) $R-T$ behavior of the CoSi$_2$ film (with a peak near the transition temperature) measured at two different currents at $\mu_0H$ = 0 T. (b) Magnetic field variation of resistance ($R-\mu_0H$) of the film measured at various temperatures ranging from 350 mK to 1 K with 0.1 mA current. (c) Optical microscope image of the sample with contact pads in van der Pauw geometry. [The current and voltage probes were connected mutually perpendicular to each other (along the diagonal) as marked by arrows for Hall effect measurement]. (d) Normal state Hall resistance of the CoSi$_2$ film measured at $T$ = 1 K. The carrier density ($n$) and Hall coefficient ($R_{\rm{H}}$) are mentioned inside the figure for convenience.}
\end{figure*}
The temperature dependence of the resistance ($R-T$) of the CoSi$_2$ film measured in van-der Pauw technique for two different currents under zero magnetic field is presented in Fig. \ref{RT_RH_CoSi2} (a). Firstly, we observe a superconducting transition around 900 mK with a prominent peak near the transition temperature. Secondly, although there is almost an order of magnitude drop in resistance in the superconducting state with respect to the normal state, the resistance value in the superconducting state is still non-zero. This again hints towards the Josephson coupling between the superconducting grains, which we will discuss later.

\begin{figure*}[h]
\centering
\includegraphics[width=0.95\textwidth]{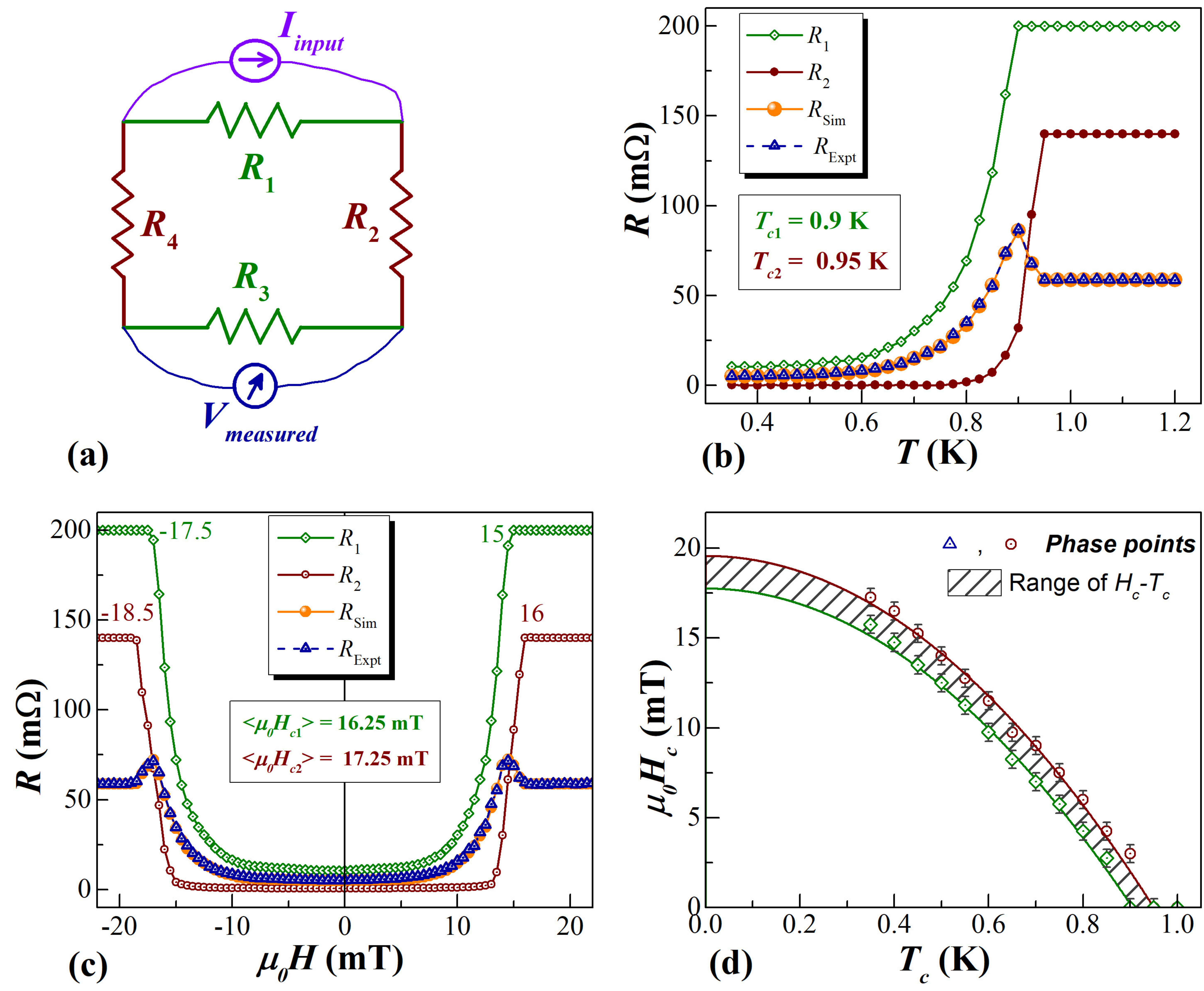}
\caption{\label{RT_model} (a) Schematic of effective resistance model to account for the peak near the superconducting transition. Simulation using the effective two-resistance model of $-$ (b) experimental $R$ - $T$ data at zero field, and (c) experimental $R$ - $\mu_0H$ data at 350 mK. (d) $\mu_0 H_c$ - $T_c$ phase diagram generated from the experimentally measured $R-H$ curves, obtained using the effective two-resistance model. The solid lines indicate fitting using the BCS model. The shaded portion shows the broadened phase boundary between the normal and the superconducting state of the total film.}
\end{figure*}

The peak in the resistive transition is easily reproduced by considering simultaneously the inhomogeneity of the sample and the geometry of the electrical probes. Assuming a simple equivalent circuit representation with normal state resistances $R_1$, $R_2$, $R_3$, and $R_4$ between each couple of probes in the vdP configuration as in Fig. \ref{RT_model}(a), the measured resistance $R_{meas}$ = $V_{measured}$/$I_{input}$ is given by \cite{vaglio1993explanation, park1997resistance}
\begin{equation}\label{Eq:2 R model}
R_{meas} = \frac{R_1 \times R_3}{ \sum_{i} R_i}
\end{equation}
Because of the inhomogeneity of our sample, the four resistors can have different $T_c$. For simplicity, we further assume that $R_1$ = $R_3$ and $R_2$ = $R_4$ at all temperatures. Then, using Eq.\ref{Eq:2 R model} we fit the zero-magnetic-field $R-T$ curve as shown in Fig. \ref{RT_model}(b).

Similar to the $R-T$ curve in Fig. \ref{RT_RH_CoSi2} (a), we notice a peak in resistance near the critical magnetic field in the $R-\mu_0H$ curve in Fig. \ref{RT_RH_CoSi2} (b). In order to explain this behaviour, it is reasonable to argue that two zones having different $T_c$, as described in previous paragraph, are also characterized by different critical magnetic fields.
To fit the $R-T$ and $R-\mu_0H$ curve we justify our arguments as follows:\\
(i) The regions with higher normal state resistance ($R_N$) behave as more inhomogeneous region with lower $T_c$ and lower $\mu_0H_c$; and conversely the regions with low $R_N$ will have higher $T_c$ and  higher $\mu_0H_c$ \cite{vaglio1993explanation, durkin2020rare},\\
(ii) Apart from the measure of $R_N$, the degree of inhomogeneity can be evaluated qualitatively from the width of the transition: a higher inhomogeneity implies a wider transition \cite{benfatto2009broadening}.

Considering the above arguments, we first fit the zero field $R-T$ curve by taking $T_c^{R_2}$ = 0.95 K $>$ $T_c^{R_1}$ = 0.9 K. Accordingly, we relate the slight $T_c$ difference of $\approx$ 50 mK to the presence of inhomogeneity/disorder \cite{kim1994new}, and thus have $\mu_0H_c^{R_2}$ $>$ $\mu_0H_c^{R_1}$,
where $\mu_0H_c^{R_{1(2)}}$ is the critical magnetic field of the zone described by the resistor $R_{1(2)}$. Using this argument, we fit the $R-H$ curve at zero temperature [see Fig. \ref{RT_model}(c)]. From these fits, we note that the two $T_c$'s are located one at the peak point of the experimental curve and the other at the point where the normal state resistance is just reached. Utilizing this observation, we extract the pair of values of $\mu_0H_c$ at different temperatures from the experimental $R-H$ curves and generate the $\mu_0H_c-T_c$ phase diagram as shown in Fig. \ref{RT_model}(d). We further note that the phase diagram does not present a sharp boundary between the normal and superconducting state, rather a broad region (shaded area in the figure), where the experimentally extracted values represent the two limits of the phase boundary. We then fit the $\mu_0H_c-T_c$ data using the BCS model (as in Eq. \ref{BCS}) and find the highest critical field $\mu_0H_c$(0) $\approx$ 20 mT, comparable to previous results on CoSi$_2$ films \cite{chiu2021observation, badoz1985superconductivity}.

We also measured the Hall effect in the CoSi$_2$ film at normal state ($T$ = 1 K) as shown in Fig. \ref{RT_RH_CoSi2}(c) and (d), which indicates that the charge carriers are holes, in consistency with previous results \cite{chiu2017ultralow, heredia2024giant}. We found the Hall coefficient $R_{\rm{H}}$ = 1.6 $\times$ 10$^{-6}$ $m^3$/C, hole concentration $n$ = 3.9 $\times$ 10$^{28}$ m$^{-3}$ and a Hall mobility $\mu_{\rm{H}}$ = 1.18 T$^{-1}$.  From this, we estimated the Fermi wave vector $k_{\rm{F}}$ [= (3$\pi^2n$)$^{1/3}$] $\approx$ 1.05 $\times$ 10$^{10}$ m$^{-1}$ and a Fermi velocity of $v_{\rm{F}}$ $\approx$ 1.22 $\times$ 10$^6$ m/s \cite{chiu2017ultralow, radermacher1993quantum, krontiras1985measurements}. We further calculated the mean free path, $l$ = 6.12 $\times$ 10$^{-10}$ m from the longitudinal resistivity ($\rho$) using the relation: $l$ = 3$\pi^2 \hbar$/($k_{\rm{F}}^2 e^2 \rho$), and estimated the disorder factor $k_{\rm{F}} l$ $\approx$ 6.43. Accordingly, the high value of $R_{\rm{H}}$ = 1.6 $\times$ 10$^{-6}$ $m^3$/C in our CoSi$_2$ film is consistent with the low value of $k_{\rm{F}} l$ in comparison with recent reports on polycrystalline  CoSi$_2$ film \cite{heredia2024giant}.
$ $\\
\textit{B. I-V characteristics}\\
\begin{figure*}[h]
\centering
\includegraphics[width=0.9\textwidth]{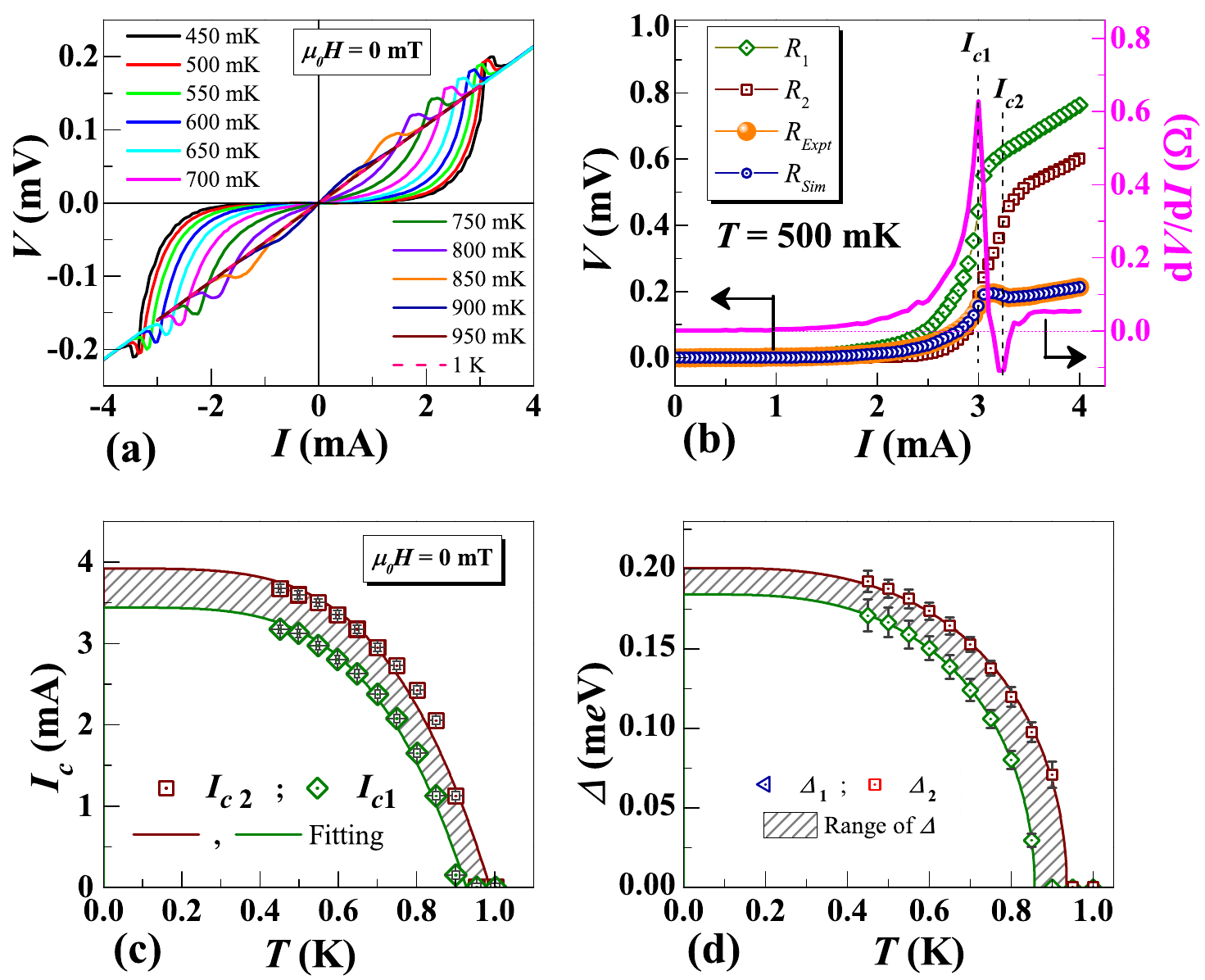}
\caption{\label{IV} (a) $I-V$ characteristic of the CoSi$_2$ film measured at various temperatures at zero magnetic field. (b) As an example, $I-V$ at 500 mK is fitted using the two-resistance model that justifies the hump in the $I-V$ curves. The corresponding critical currents $I_c1$ and $I_c2$ can be obtained from the d$V$/d$I$ curves as shown (right axis). 
(c) $I_c$-$T$ phase diagram generated from the experimentally measured $I-V$ curves.  (d) $\Delta$ vs $T$ phase diagram, where values of $\Delta$ at each temperature is estimated from the $I-V$ curves. Solid lines are fitting using simplified BCS model (see main text). The shaded portions in both panels (c) and (d) indicate the broadened phase boundary between the normal and the superconducting state of the total film.}
\end{figure*}
To find the critical current ($I_c$), we measured the $I-V$ characteristics of the CoSi$_2$ film at various temperatures under zero magnetic field, as shown in Fig. \ref{IV}(a). Instead of a sharp transition to a normal state (recognised by linear region), we observe exponential type of behaviour, which probably is due to presence of inhomogeneity \cite{venditti2019nonlinear, veyrat2023berezinskii}. The linear region above a certain (critical) current indicates the normal state region. Interestingly we also observe a peculiar hump near the critical current. By considering the equivalent resistor circuit as in Fig. \ref{RT_model}(a), we can argue that the hump in $I-V$ characteristics occurs due to two different critical currents of the two different resistive regions \cite{park1997resistance} ($R_1$ and $R_2$) as shown in Fig. \ref{IV}(b) (left axis). The critical currents can be best extracted from the d$V$/d$I$ vs $I$ curves as shown in Fig. \ref{IV}(b) (right axis). From each $I-V$ curves in Fig. \ref{IV}(a), we determine the corresponding $I_{c1}$ and $I_{c2}$ values and thus generate the $I_c-T$ phase diagram at zero magnetic field, as shown in Fig. \ref{IV}(c). From the $I-V$ characteristics in Fig. \ref{IV}(a), for each temperature, we also estimate the values of $\Delta_1$ and $\Delta_2$ (in m$e$V units) (where 2$\Delta$ = $E_g$, the energy gap) by taking the voltage values corresponding to $I_{c1}$ and $I_{c2}$ and generate the $\Delta - T$ phase diagram which is shown in Fig. \ref{IV}(d).

Next, in order to speculate the possibility of internal Josephson coupling from the $I_c-T$ curve, we take into account the Ambegaokar-Baratoff (AB) relation \cite{venditti2019nonlinear, ambegaokar1963tunneling}:
\begin{equation}\label{AB_Ic-Tc}
I_c (T) = I_c(0) \times \frac{\Delta(T)}{\Delta(0)} \times tanh\left(\frac{\Delta(T)}{2k_{\rm{B}}T}\right)
\end{equation}
where $\Delta(T)$ is the temperature dependent energy gap. In order to do so, we first estimate the zero temperature energy gap, $\Delta(0)$, for each region by fitting the $\Delta(T)$ vs $T$ curves (Fig. \ref{IV}d) using an approximate formula that reproduces well the BCS
behavior \cite{venditti2019nonlinear}:
\begin{equation}\label{Modified_BCS_gap}
\Delta_i(T) = \Delta_i(0)\left[1-\frac{1}{3}\left(\frac{T}{T_c^i}\right)^4\right]\sqrt{1-\left(\frac{T}{T_c^i}\right)^4}
\end{equation}
We obtained $\Delta_1(0)$ = 0.185(2) m$e$V and $\Delta_2(0)$ = 0.199(3) m$e$V; however, these are somewhat higher when compared with $\Delta$(0) values calculated using the BCS prediction \cite{tinkham2004introduction, bardeen1957theory}: $\Delta$(0)=1.76$k_{\rm{B}}T_c$ for clean systems. This kind of situations have been encountered earlier due to the presence of inhomogeineity \cite{giaever1961study, douglass1964energy}. Nonetheless, utilizing these in Equation \ref{AB_Ic-Tc}, we fit the $I_c-T$ curve to obtain the zero temperature critical currents $I_{c1}$(0) = 3.44(6) mA  and $I_{c2}$(0) = 3.93(9) mA. The fitting of the $I_c-T$ curves with Ambegaokar-Baratoff formula indicates the possibility of the presence of weak Josephson coupling between the polycrystalline regions, and is still under our scrutiny. 
\section{Conclusion}
We have presented the conversion of one superconducting system (self-organized HDNM Cobalt film on Si) into another (CoSi$_2$ film) by annealing the former under vacuum and compared their superconducting properties. The HDNM Co film, which had $T_c \approx $ 5K  showed anisotropy in the upper critical field with respect to its in-plane and out-of-plane directions, comparable to the HDNM Co grain dimensions. CoSi$_2$ films showed $T_c$ around 0.9 K $-$ somewhat less compared to its epitaxial form. However, the critical current densities and upper critical fields were comparable with its previously reported results. Measurement performed in van-der Pauw geometry revealed some distinctive features near the transition temperature (also field and current) which were explained considering multiple-resistance model for polycrystalline regions (confirmed by HRTEM) of varying superconducting properties. Analysis of the critical current and energy gap parameter revealed possible presence of Josephson coupling between the polycrystalline CoSi$_2$ regions. Superconducting materials with internal Josephson junctions, \textit{e.g.}, disordered superconductors or granular superconductors can have high kinetic inductance. While normal aluminium is commonly used for fabricating transmon qubits, granular aluminium superconductors, with their high kinetic inductance, are used in fabricating fluxonium qubits \cite{grunhaupt2019granular, winkel2020implementation}, desirable for building quantum annealing computers. While the Co thin film superconductor is a disordered superconductor, recent studies suggest that CoSi$_2$ is a valuable addition to the toolkit of materials for quantum circuit fabrication \cite{mukhanova2024kinetic}.

We further aim to develop both polycrystalline and epitaxial CoSi$_2$ films by annealing such trilayer structured Co films, for fabricating superconducting quantum circuits in the near future. CoSi$_2$, being resistant to oxidation, may also be valuable for the suppression of undesirable two-level-systems.

\section*{Acknowledgments}
S.M, B.B and B.N.D acknowledge TCG CREST for funding. S.M and B.B also acknowledge help from Prof. Biswajit Karmakar of SINP during measurement in dilution refrigerator and Mr. Arnab Bhattacharya and Mr. Afsar Ahmed of SINP during measurement in PPMS.  

\section*{References}


\begin{thebibliography}{10}
\expandafter\ifx\csname url\endcsname\relax
  \def\url#1{{\tt #1}}\fi
\expandafter\ifx\csname urlprefix\endcsname\relax\def\urlprefix{URL }\fi
\providecommand{\eprint}[2][]{\url{#2}}

\bibitem{kjaergaard2020superconducting}
Kjaergaard M, Schwartz M~E, Braum{\"u}ller J, Krantz P, Wang J~I~J, Gustavsson
  S and Oliver W~D 2020 {\em Annu. Rev. Condens. Matter Phys.\/} {\bf 11}
  369--395

\bibitem{lau2022nisq}
Lau J~W~Z, Lim K~H, Shrotriya H and Kwek L~C 2022 {\em AAPPS Bulletin\/} {\bf
  32} 27

\bibitem{nakamura1999coherent}
Nakamura Y, Pashkin Y~A and Tsai J 1999 {\em Nature\/} {\bf 398} 786--788

\bibitem{siddiqi2021engineering}
Siddiqi I 2021 {\em Nat. Rev. Mater.\/} {\bf 6} 875--891

\bibitem{chiu2017ultralow}
Chiu S~P, Yeh S~S, Chiou C~J, Chou Y~C, Lin J~J and Tsuei C~C 2017 {\em ACS
  Nano\/} {\bf 11} 516--525

\bibitem{zhang2003metal}
Zhang S~L and {\"O}stling M 2003 {\em Crit. Rev. Solid State Mater. Sci.\/}
  {\bf 28} 1--129

\bibitem{furukawa1983epitaxial}
Furukawa S and Ishiwara H 1983 {\em Jpn. J. Appl. Phys.\/} {\bf 22} 21

\bibitem{shi2000growth}
Shi J, Irie T, Takahashi F and Hashimoto M 2000 {\em Thin Solid Films\/} {\bf
  375} 37--41

\bibitem{mahato2012nanodot}
Mahato J~C, Das D, Juluri R~R, Batabyal R, Roy A, Satyam P~V and Dev B~N 2012
  {\em Appl. Phys. Lett.\/} {\bf 100}

\bibitem{tung1982growth}
Tung R, Bean J, Gibson J, Poate J and Jacobson D 1982 {\em Appl. Phys. Lett.\/}
  {\bf 40} 684--686

\bibitem{tsutsumi1997superconductivity}
Tsutsumi K, Takayanagi S and Hirano T 1997 {\em Physica B\/} {\bf 237} 310--311

\bibitem{kittel2018introduction}
Kittel C and McEuen P 2018 {\em {Introduction to Solid State Physics}\/} 8th ed
  (John Wiley \& Sons) ISBN 9780471415268, 047141526X, 0471680575

\bibitem{shimizu2001superconductivity}
Shimizu K, Kimura T, Furomoto S, Takeda K, Kontani K, Onuki Y and Amaya K 2001
  {\em Nature\/} {\bf 412} 316--318

\bibitem{banu2020inhomogeneous}
Banu N, Aslam M, Paul A, Banik S, Das S, Datta S, Roy A, Das I, Sheet G,
  Waghmare U {\em et~al.\/} 2020 {\em Europhys. Lett.\/} {\bf 131} 47001 ; 2017 {\em arXiv preprint arXiv:1710.06114\/}


\bibitem{banu2017evidence}
Banu N, Singh S, Satpati B, Roy A, Basu S, Chakraborty P, Movva H~C, Lauter V
  and Dev B 2017 {\em Sci. Rep.\/} {\bf 7} 1--8

\bibitem{banu2018high}
Banu N, Singh S, Basu S, Roy A, Movva H~C, Lauter V, Satpati B and Dev B 2018
  {\em Nanotechnology\/} {\bf 29} 195703

\bibitem{yoo2000new}
Yoo C, Cynn H, S{\"o}derlind P and Iota V 2000 {\em Phys. Rev. Lett.\/} {\bf
  84} 4132

\bibitem{baek2014hybrid}
Baek B, Rippard W~H, Benz S~P, Russek S~E and Dresselhaus P~D 2014 {\em Nat.
  Commun.\/} {\bf 5} 3888

\bibitem{bhatia2022aspects}
Bhatia E and Senapati K 2022 {\em Supercond. Sci. Technol.\/} {\bf 35} 094004

\bibitem{pompeo2014thermodynamic}
Pompeo N, Torokhtii K, Cirillo C, Samokhvalov A, Ilyina E, Attanasio C, Buzdin
  A~I and Silva E 2014 {\em Phys. Rev. B\/} {\bf 90} 064510

\bibitem{yamashita2005superconducting}
Yamashita T, Tanikawa K, Takahashi S and Maekawa S 2005 {\em Phys. Rev.
  Lett.\/} {\bf 95} 097001

\bibitem{gyenis2021experimental}
Gyenis A, Mundada P~S, Di~Paolo A, Hazard T~M, You X, Schuster D~I, Koch J,
  Blais A and Houck A~A 2021 {\em PRX Quantum\/} {\bf 2} 010339

\bibitem{lee2006multiple}
Lee P~S, Pey K~L, Chow F, Tang L, Tung C~H, Wang X and Lim G 2006 {\em IEEE
  Electron Device Letters\/} {\bf 27} 237--239

\bibitem{buckel2008superconductivity}
Buckel W and Kleiner R 2008 {\em {Superconductivity: fundamentals and
  applications}\/} (John Wiley \& Sons)

\bibitem{adkins1980increased}
Adkins C, Thomas J and Young M 1980 {\em J. Phys. C: Solid State Phys.\/} {\bf
  13} 3427

\bibitem{durkin2020rare}
Durkin M, Garrido-Menacho R, Gopalakrishnan S, Jaggi N~K, Kwon J~H, Zuo J~M and
  Mason N 2020 {\em Phys. Rev. B\/} {\bf 101} 035409

\bibitem{abeles1977effect}
Abeles B 1977 {\em Phys. Rev. B\/} {\bf 15} 2828

\bibitem{dynes1978two}
Dynes R, Garno J and Rowell J 1978 {\em Phys. Rev. Lett.\/} {\bf 40} 479

\bibitem{helfand1966temperature}
Helfand E and Werthamer N 1966 {\em Phys. Rev.\/} {\bf 147} 288

\bibitem{werthamer1966temperature}
Werthamer N, Helfand E and Hohenberg P 1966 {\em Phys. Rev.\/} {\bf 147} 295

\bibitem{chandrasekhar1962note}
Chandrasekhar B 1962 {\em Appl. Phys. Lett.\/} {\bf 1} 7--8

\bibitem{clogston1962upper}
Clogston A~M 1962 {\em Phys. Rev. Lett.\/} {\bf 9} 266

\bibitem{vaglio1993explanation}
Vaglio R, Attanasio C, Maritato L and Ruosi A 1993 {\em Phys. Rev. B\/} {\bf
  47} 15302

\bibitem{park1997resistance}
Park M, Isaacson M and Parpia J 1997 {\em Phys. Rev. B.\/} {\bf 55} 9067

\bibitem{benfatto2009broadening}
Benfatto L, Castellani C and Giamarchi T 2009 {\em Phys. Rev. B\/} {\bf 80}
  214506

\bibitem{kim1994new}
Kim J~J, Kim J, Shin H~J, Lee H~J, Lee S, Park K~W and Lee E~H 1994 {\em J.
  Phys.: Condens. Matter\/} {\bf 6} 7055

\bibitem{chiu2021observation}
Chiu S~P, Tsuei C, Yeh S~S, Zhang F~C, Kirchner S and Lin J~J 2021 {\em Sci.
  Adv.\/} {\bf 7} eabg6569

\bibitem{badoz1985superconductivity}
Badoz P, Briggs A, Rosencher E and d'Avitaya F~A 1985 {\em J. Physique Lett.\/}
  {\bf 46} 979--983

\bibitem{heredia2024giant}
Heredia E~A, Chiu S~P, Nguyen B~A~V, Wang R~T, Wu C~Y, Yeh S~S and Lin J~J 2024
  {\em Phys. Rev. B\/} {\bf 110} 024201

\bibitem{radermacher1993quantum}
Radermacher K, Monroe D, White A~E, Short K~T and Jebasinski R 1993 {\em Phys.
  Rev. B\/} {\bf 48} 8002

\bibitem{krontiras1985measurements}
Krontiras C, Salmi J, Gr{\"o}nberg L, Suni I, Heleskivi J and Rissanen A 1985
  {\em Thin Solid Films\/} {\bf 125} 93--99

\bibitem{venditti2019nonlinear}
Venditti G, Biscaras J, Hurand S, Bergeal N, Lesueur J, Dogra A, Budhani R,
  Mondal M, Jesudasan J, Raychaudhuri P {\em et~al.\/} 2019 {\em Phys. Rev.
  B\/} {\bf 100} 064506

\bibitem{veyrat2023berezinskii}
Veyrat A, Labracherie V, Bashlakov D~L, Caglieris F, Facio J~I, Shipunov G,
  Charvin T, Acharya R, Naidyuk Y, Giraud R {\em et~al.\/} 2023 {\em Nano
  Letters\/} {\bf 23} 1229--1235

\bibitem{ambegaokar1963tunneling}
Ambegaokar V and Baratoff A 1963 {\em Phys. Rev. Lett.\/} {\bf 10} 486

\bibitem{tinkham2004introduction}
Tinkham M 2004 {\em {Introduction to Superconductivity}\/} vol~1 (Dover,
  Mineola (New York))

\bibitem{bardeen1957theory}
Bardeen J, Cooper L~N and Schrieffer J~R 1957 {\em Phys. Rev.\/} {\bf 108} 1175

\bibitem{giaever1961study}
Giaever I and Megerle K 1961 {\em Phys. Rev.\/} {\bf 122} 1101

\bibitem{douglass1964energy}
Douglass~Jr D and Meservey R 1964 {\em Phys. Rev.\/} {\bf 135} A19

\bibitem{grunhaupt2019granular}
Gr{\"u}nhaupt L, Spiecker M, Gusenkova D, Maleeva N, Skacel S~T, Takmakov I,
  Valenti F, Winkel P, Rotzinger H, Wernsdorfer W {\em et~al.\/} 2019 {\em Nat.
  Mater.\/} {\bf 18} 816--819

\bibitem{winkel2020implementation}
Winkel P, Borisov K, Gr{\"u}nhaupt L, Rieger D, Spiecker M, Valenti F, Ustinov
  A~V, Wernsdorfer W and Pop I~M 2020 {\em Phys. Rev. X\/} {\bf 10} 031032

\bibitem{mukhanova2024kinetic}
Mukhanova E, Zeng W, Heredia E~A, Wu C~W, Lilja I, Lin J~J, Yeh S~S and Hakonen
  P 2024 {\em APL Materials\/} {\bf 12}

\end{thebibliography}
\providecommand{\newblock}{}

\end{document}